\documentclass[aps,prl,reprint,superscriptaddress,showkeys]{revtex4-2}

\usepackage{graphicx}
\usepackage{xcolor}
\usepackage{amsmath}

\begin{document}

% Use the \preprint command to place your local institutional report
% number in the upper righthand corner of the title page in preprint mode.
% Multiple \preprint commands are allowed.
% Use the 'preprintnumbers' class option to override journal defaults
% to display numbers if necessary
%\preprint{}

%Title of paper
\title{Structural analysis and the sum of nodes' betweenness centrality in complex networks}

% repeat the \author .. \affiliation  etc. as needed
% \email, \thanks, \homepage, \altaffiliation all apply to the current
% author. Explanatory text should go in the [] 's, actual e-mail
% address or url should go in the {} 's for \email and \homepage.
% Please use the appropriate macro foreach each type of information

% \affiliation command applies to all authors since the last
% \affiliation command. The \affiliation command should follow the
% other information
% \affiliation can be followed by \email, \homepage, \thanks as well.
\author{Ronghao Deng}
%\email[]{llimeizhu@163.com}
%\homepage[]{Your web page}
%\thanks{}
%\altaffiliation{}
\affiliation{School of Science, Jiangsu University of Science and Technology, Zhenjiang 212100, China}

\author{Meizhu Li}
%\email[]{llimeizhu@163.com}
%\homepage[]{Your web page}
%\thanks{}
%\altaffiliation{}
\affiliation{School of Computer Science and Communication Engineering, Jiangsu University, Zhenjiang, Jiangsu, China}

\author{Qi Zhang}
\email[]{qi.zhang@just.edu.cn}
%\homepage[]{Your web page}
%\thanks{}
%\altaffiliation{}
\affiliation{School of Science, Jiangsu University of Science and Technology, Zhenjiang 212100, China}
\affiliation{Lorentz Institute for Theoretical Physics, Leiden University, PO Box 9504, 2300 RA Leiden, The Netherlands}

%Collaboration name if desired (requires use of superscriptaddress
%option in \documentclass). \noaffiliation is required (may also be
%used with the \author command).
%\collaboration can be followed by \email, \homepage, \thanks as well.
%\collaboration{}
%\noaffiliation
\date{\today}

\begin{abstract}
Structural analysis in network science is finding the information hidden from the topology structure of complex networks. Many methods have already been proposed in the research on the structural analysis of complex networks to find the different structural information of networks. In this work, the sum of nodes' betweenness centrality (SBC) is used as a new structural index to check how the structure of the complex networks changes in the process of the network's growth. 
We build two four different processes of network growth to check how the structure change will be manifested by the SBC. We find that when the networks are under Barabási–Albert rule, the value of SBC for each network grows like a logarithmic function. However, when the rule that guides the network's growth is the Erdős–Rényi rule, the value of SBC will converge to a fixed value. It means the rules that guide the network's growth can be illustrated by the change of the SBC in the process of the network's growth. In other words, in the structure analysis of complex networks, the sum of nodes' betweenness centrality can be used as an index to check what kinds of rules guide the network's growth. 
\end{abstract}

% insert suggested PACS numbers in braces on next line
% insert suggested keywords - APS authors don't need to do this
%\keywords{}

%\maketitle must follow title, authors, abstract, \pacs, and \keywords
\maketitle

% body of paper here - Use proper section commands
% References should be done using the \cite, \ref, and \label commands

\section{Introduction}
The birth of complex networks can be attributed to the finding of scale-free and small-world properties in complex systems at the beginning of this century~\cite{watts1998collective,barabasi2003scale}. The mathematical definition of complex networks is based on the graph, which is comprised of sets of nodes and edges. The application of the complex networks can be found in the research of physics\cite{newman2004analysis,garlaschelli2009generalized,bianconi2001bose}, biology\cite{von2002comparative,korcsmaros2017next,gosak2018network,shai2017multilayer}, neuroscience\cite{schneidman2006weak,bassett2017network}, financial systems~\cite{battiston2016complexity,almog2014binary,kenett2015network}, and even social sciences~\cite{bello2016social,jusup2022social}. These new network models are widely used to describe those new phenomena and behaviours in complex systems. Complex networks not only provide new tools for problems existing in different research fields. Moreover, some new phenomena related to complex networks have been found in the basic research of physics~\cite{cimini2019statistical,battiston2020networks}. For instance, the breaking of ensemble equivalence in statistical physics~\cite{squartini2015breaking,barvinok2012matrices,hollander2018breaking,squartini2017reconnecting,zhang2022strong}.

Complex networks are characterized by a simple underlying philosophical logic. In these networks, nodes represent the units within a system, while the edges between nodes depict the interconnections and relationships among these units~\cite{cimini2019statistical,newman2004analysis}. Thus, the main problem in the research on the networks' structure analysis is how to find the information hidden in the topology structure of those networks. Many methods have been proposed to achieve this intent in the research on network science, such as the nodes' centrality, the network's structural entropy, the detection of the community structure in networks, and so on. 

Among those methods of structure analysis, there is a special kind of method, which is called the nodes' structural characteristic sum. For instance, the sum of the nodes' centrality, which is the sum of each node's degree. It is also called the total degree in the network, and it can be used to check the density of the structure of networks, which is a very important method in the research on network structure analysis. The definition of total degree inspired us to think that the sum of different kinds of node centrality may give us a new kind of index of the network's structure analysis. For instance, the sum of each node's betweenness centrality (SBC). As we know, the betweenness is a global structure characteristic of the nodes in the networks, which is totally different from the node's degree~\cite{prountzos2013betweenness,lee2012qube,brandes2001faster}. This also means the SBC can give us unique information about the network's topology structure~\cite{barthelemy2004betweenness,newman2005measure,leydesdorff2007betweenness}. 

In this work, in order to check what is the unique information that the SBC can give us, we generated several networks according to two different growth rules: the Barabási–Albert rule and the the Erdős–Rényi rule.  We find that when the networks are under Barabási–Albert rule, the value of SBC for each network grows like a logarithmic function. However, when the rule that guides the network's growth is the Erdős–Rényi rule, the value of SBC will converge to a fixed value. 
We also find that the randomness inherent in edge generation can lead to small fluctuations in SBC. To mitigate this, we conducted multiple experiments to calculate the mean and reduce randomness, yet significant fluctuations persisted. Hence, we proceeded to extract the network structure and compare it with SBC, uncovering the relationship between SBC and network connectivity. Those results show that SBC can be used to check what kind of growth rules the network's growth follows.  

In order to verify our findings on the properties of SBC, we have decomposed the US-airline network by the k-shell method and calculated the SBC for the networks in the process of decomposition. We find that based on the change of SBC of the US-airline network under k-shell, it is BA networks. We also decomposed an ER network by the k-shell, and the value change of SBC in that process also shows us it is an ER-ruled based network.  

The rest of this paper is organised as follows. In section 2, the definition of the sum of nodes' betweenness centrality (SBC) is proposed. In section 3, the relationship between SBC and networks' growth is illustrated. In section 4, we use k-shell to decompose those real networks and check how the SBC will change in that process. Conclusions are given in section 5.

\section{Definition of the sum of nodes' betweenness centrality}
The computation of the sum of nodes' betweenness centrality relies on the concept of betweenness centrality itself. For a graph $\mathbf{G}(\mathcal{N},\mathcal{E})$, where $\mathcal{N}$ denotes the set of nodes and $\mathcal{E}$ the set of edges, the betweenness centrality of each node is determined by the count of shortest paths passing through that node for all pairs of distinct nodes. Specifically, the betweenness centrality of node 
$i$ is formally defined as:

\begin{equation}
{BC}(i)=\sum_{s\neq i\neq t}\frac{\sigma_{st}(i)}{\sigma_{st}}.
\label{eq_betweenness_definition}
\end{equation}

where $\sigma_{st}$ represents the total number of shortest paths from node $s$ to node $t$,and $\sigma_{st}(i)$ denotes the number of those paths that traverse node $i$.

In this study, our attention is confined to undirected networks. Thus, the betweenness centrality ($BC(i)$) of node $i$ is computed as the ratio of the total number of shortest paths passing through node $i$ to the total number of pairs of nodes in the network, which is $\frac{n(n-1)}{2}$. Therefore, the betweenness centrality ($BC(i)$) value for node $i$ falls within the range of $[0,1]$.

Utilizing the betweenness centrality of each node, we can aggregate these values to derive the definition of the sum of nodes' betweenness centrality in complex networks, denoted as SBC.

\begin{equation}
     SBC=\sum_{i=1}^n{BC}(i).
\end{equation}
\\

The symbol $n$ denotes the total number of nodes in the network $\mathbf{G}$. As the network expands, the value of $n$ naturally increases, leading to changes in the betweenness centrality of all nodes, which may remain constant in some cases. Consequently, the sum of nodes' betweenness centrality (SBC) must also vary. Adding new nodes influences the overall network structure, resulting in significant and minor alterations. Thus, the SBC serves as a meaningful indicator of changes in the network structure to a certain extent.

\section{The relationship between SBC and networks' growth}

The sum of nodes' betweenness centrality (SBC) in a complex network is determined by the betweenness centrality values of its nodes and the total node count. The network's structure directly influences the betweenness centrality of each node. In this section, we aim to investigate the relationship between SBC and network structure. To accomplish this, we will simulate network growth from a single node using two distinct growth rules. Through this analysis, we will elucidate the factors driving changes in SBC and highlight the variations in SBC resulting from different network structures.

The development of network structure during the growth process is regulated by growth rules, which determine how newly added nodes influence the network structure and subsequently affect the betweenness centrality of each node.  Therefore, before investigating the correlation between SBC and network structure, it is crucial to introduce two distinct network growth rules.

\subsection{The growth of the network}
Network growth entails initially adding a new node to the network and subsequently attempting to connect this new node with all existing nodes. The probability of a connection is determined by the growth rules governing the network.

This paper explores network growth from a single node using various growth rules. To investigate the differences in network structure represented by SBC, we employ two distinct growth rules to generate different network structures. This section outlines these two growth rules, both derived from network growth models proposed by different researchers~\cite{barabasi2003scale,erdHos1961strength}.

The first rule is based on the Barab{\'a}si-Albert (BA) model~\cite{barabasi2003scale}, a classic algorithm for generating random scale-free networks in network science. Many real-world networks, such as the World Wide Web, the Internet, and social networks like Facebook and Twitter, exhibit this scale-free feature. In these real networks, hub nodes are prevalent, and the connections to these hubs increase as the network expands. Therefore, a rule based on the BA model is essential to guide network growth, which we will henceforth refer to as the BA rule.

When the network grows according to the BA rule, newly added nodes are more likely to link with hub nodes than with marginal nodes. This probability is determined by both the degree of the target node and the total degree in the network. For instance, if the degree of node $i$ is denoted as $d(i)$, then the probability for a newly added node to connect with node $i$ in the seed network is given by

\begin{equation}
P_{\textrm{BA}}(i) = \frac{d(i)}{\sum_{i=1}^{\tilde{n}}d(i)}
\end{equation}

Here, $\tilde{n}$ represents the temporary size of the network. As the network grows, the value of $\tilde{n}$ increases. The definition of $P_{\textrm{BA}}(i)$ clearly illustrates that nodes with higher degrees have a greater probability of connecting with the newly added node when the network grows under the BA rule.

Following the BA network growth rule, we initiate with a one-node network, and with each new node added, it randomly connects based on the priority of existing node degrees. Furthermore, only one link is generated each time a new node is introduced.

The second rule is based on the Erd{\H{o}}s-R{\'e}nyi (ER) model~\cite{erdHos1961strength}. The ER model is a classical algorithm used in random network generation. Under this model, each new node added has an equal probability of connecting to every node in the initial network. Therefore, following the ER rule, the new node added during the network's growth also has an equal probability of linking to all nodes in the initial network. The value of this probability is determined by the total number of nodes in the initial network. For instance, the probability of the new node connecting to node $i$ is given by:

\begin{equation}
P_{\textrm{ER}}(i) = \frac{1}{\tilde{n}},
\end{equation}

where $\tilde{n}$ still represents the temporary size of the network. Obviously, with the growth of the network, the number of new nodes in the network forming links with existing nodes decreases with the increase of n.
Therefore, when the ER rule is applied during the network's growth, the network structure tends towards homogeneity.

Due to the varying structural priorities governing connections between new and existing nodes, these two rules lead to the growth of distinct network topologies. Networks evolving under the BA growth rule exhibit structural heterogeneity as new nodes preferentially attach to high-degree nodes. Conversely, networks following the ER growth rule demonstrate structural homogeneity, as the connection probability for new nodes equals that of existing nodes, resulting in minimal degree disparities across the network. The homogeneity of the network structure can be assessed through degree distribution and the SBC proposed in this paper.

\subsection{Changes in the sum of nodes' betweenness centrality}
We employ two network growth rules, BA and ER, to expand multiple networks and observe the evolution of the SBC as the networks develop. We note significant disparities in the SBC evolution between networks grown under the BA and ER rules, attributing these distinctions to changes in the contact network's structure. Furthermore, networks grown under the ER rule present a unique case, which we also elucidate in the context of network structure.

\subsubsection{The sum of nodes' betweenness centrality under BA rule}

\begin{figure}[htbp]
    \includegraphics[width=8.6cm]{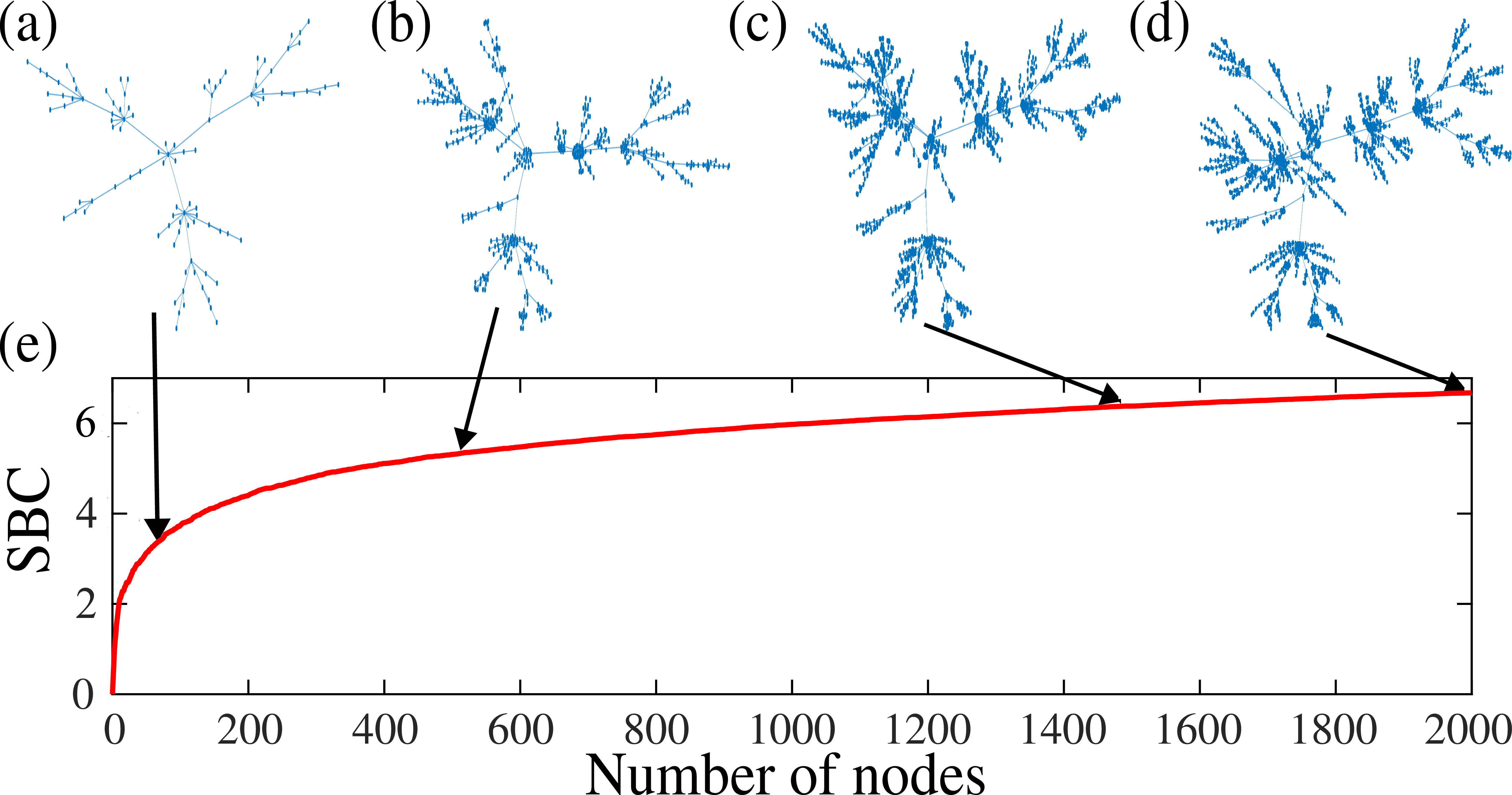}
\caption{The curve depicted in the figure illustrates the trend in SBC changes for networks evolving under the BA rule. It is evident that the SBC evolution in networks following the BA rule resembles that of an exponential function. Initially, SBC experiences rapid growth, which gradually decelerates as additional nodes are incorporated.   To investigate the underlying reasons, we extracted four network structure diagrams representing various node quantities corresponding to the arrows in the figure. }

\label{BA}
\end{figure}

The SBC of networks evolving under the BA rule undergoes a noticeable increase over time. Networks adhering to BA rules exhibit a core-periphery structure, wherein a small number of nodes possess high degrees of betweenness centrality. As the network expands, the betweenness centrality of these nodes escalates, while most nodes maintain a degree of one. Consequently, the evolution of SBC during network growth manifests significant changes. We illustrate in Fig.\ref{BA} the structural evolution of the network grown under the BA rule, specifically highlighting the change in SBC.

We noticed that initially, core nodes demonstrate significant betweenness centrality, driving rapid growth in SBC due to the constrained total number of node pairs.   However, as the network expands following the BA growth rule, the majority of new nodes consistently form connections with several core nodes.   As a result, the betweenness centrality of these core nodes fails to keep up with the escalating total number of node pairs, leading to a deceleration in growth rate.
\subsubsection{The sum of nodes' betweenness centrality under ER rule}

\begin{figure}[htbp]
    \includegraphics[width=8.6cm]{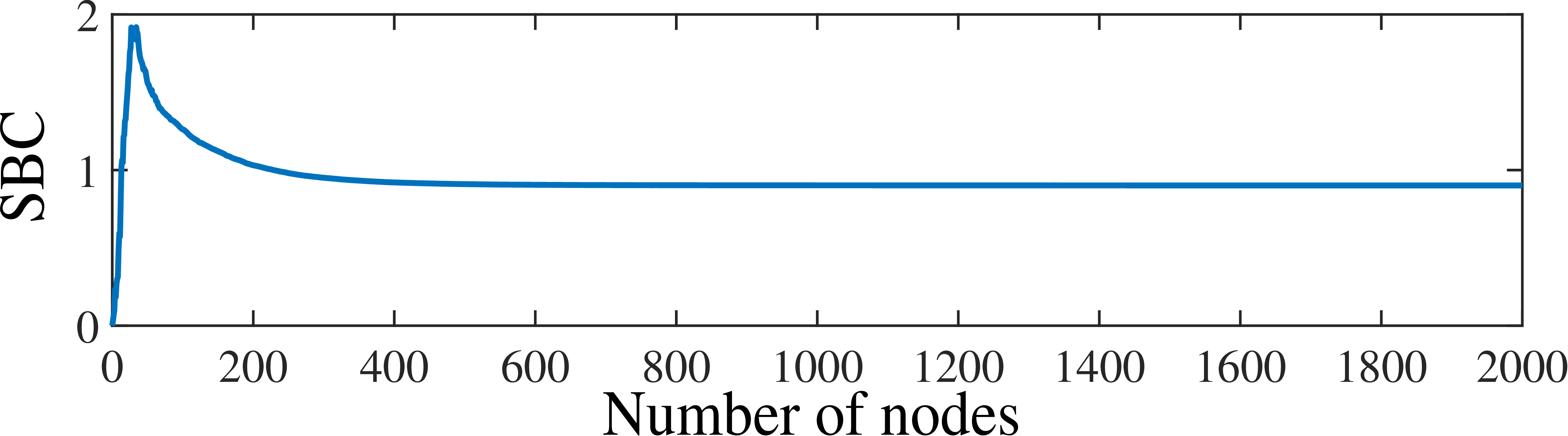}
\caption{The curve in the figure shows the change of network SBC growing under the ER rule. We can see that the growth network using ER rules will have a peak value in the initial stage, and with the increase of network size, SBC will gradually decrease and become stable. }

\label{ER}
\end{figure}

There is little disparity between the degree and betweenness centrality of each node in the network developed under ER rules, indicating a homogeneous network structure. Additionally, we expand a network comprising two thousand nodes to examine the impact on the SBC of a network cultivated using ER rules. We show in Fig.\ref{ER} how SBC changes when growing a network using ER rules.

We will explain why the spike occurs below. A network grown using ER rules has many links per node, and the shortest path between points is very small, so the betweenness centrality of each node is small and the value is small. After the network has a certain scale, adding a new node will not cause a large change in SBC. We can also take this to an extreme and consider a completely connected network (points linked to each other), such that every node has 0 betweenness centrality and SBC is 0, even if a new node is added, it has 0 betweenness centrality and SBC is stable at 0. We can also use this example to explain why the SBC of an ER regular-grown network is stable.
\subsubsection{Sum of nodes' betweenness centrality reflects the rules of network growth}
We know that the networks grown by two different growth rules have different structural characteristics, and we can use many indicators such as degree distribution to show the differences in network structure. Although we show in Fig.\ref{BA} and Fig.\ref{ER} that the change of SBC of the network guided by these two growth rules is very different, in order to treat the index SBC more rigorously and to have a deeper understanding of the changes in the structural characteristics of the network, we need to change the growth rules during the growth of the network and observe whether SBC can reflect the change of the network structure. We first grew a network of 500 nodes using the BA rule, and then grew it to 1,000 nodes using the ER rule. An example of using ER rule first and then BA rule is also given.

\begin{figure}[htbp]
    \includegraphics[width=8.6cm]{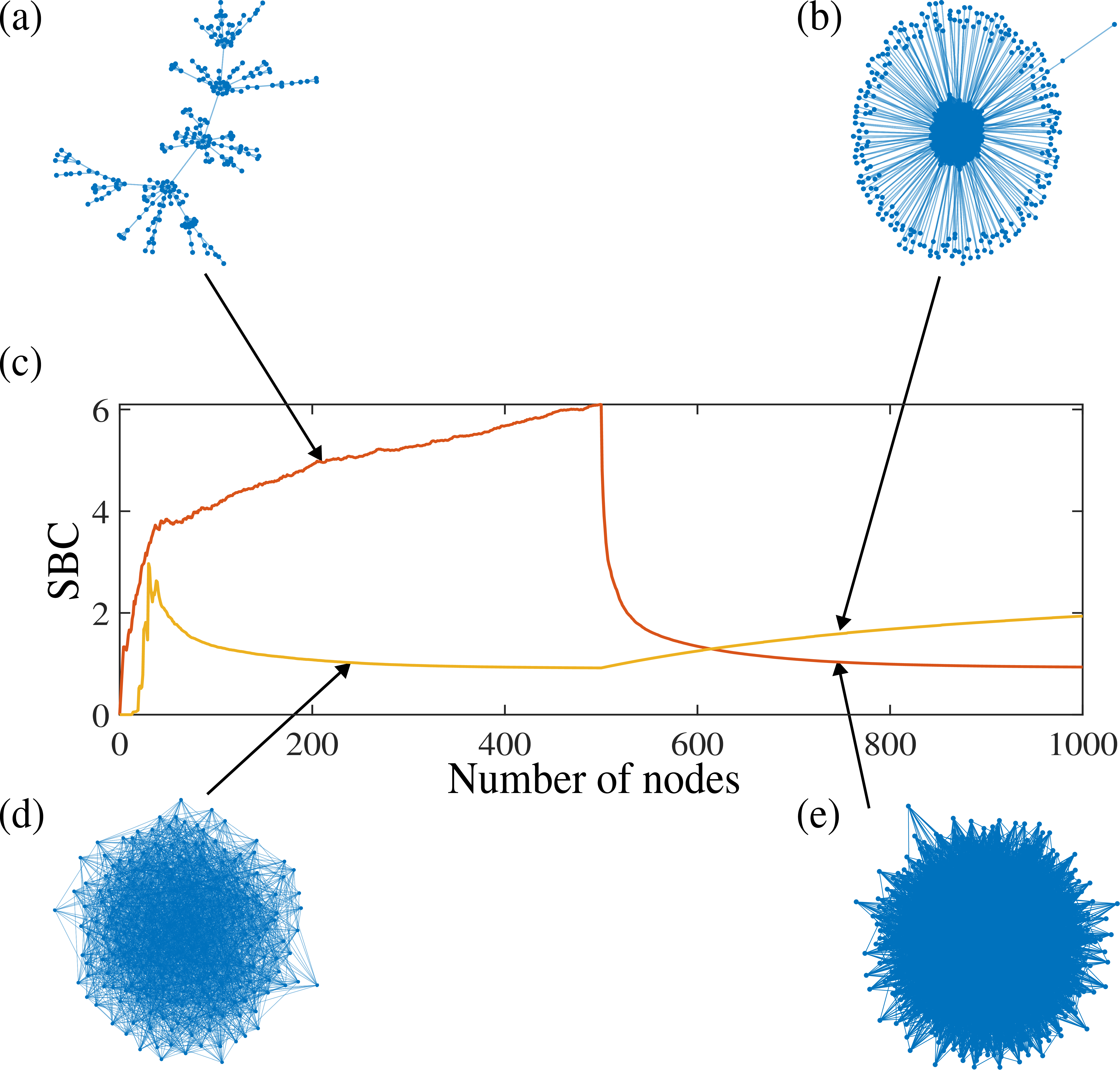}
\caption{The red line in the figure represents the change of SBC in a network that first uses the BA rule and then the ER rule. At first, SBC grows exponentially, and when it reaches 500 nodes, the ER rule makes SBC gradually decrease and eventually stabilize. The yellow line in the figure represents the change of SBC in a network that first uses ER rules and then BA rules. SBC initially peaks and then becomes stable, and SBC increases exponentially after BA rules are used. We extracted the network structure of the four stages, and the four network diagrams in the figure show the network structure of the stage pointed by the arrows.}
\label{BA(ER)-ER(BA)}
\end{figure}
It is evident that the alterations in SBC following changes in growth rules align with the trends depicted in Fig.\ref{BA} and Fig.\ref{ER}. Even when growth rules are modified after the network reaches a certain scale, SBC remains indicative of the network's structure. Moreover, compared to the traditional degree distribution, changes in SBC are more perceptible and rapid. This further underscores SBC as an index capable of reflecting both growth rules and network structure in an intuitive and reliable manner. In essence, fluctuations in SBC capture the general shifts in growth rules, while the magnitude of SBC values delineates the overall network structure.
\subsubsection{The appearance of peak value in the sum of nodes' betweenness centrality under ER rule}
To explore the underlying reasons behind the peaks observed in the SBC of networks evolving under the ER model, it is imperative to conduct a detailed examination of SBC dynamics, especially when the network size is relatively small. This is crucial because fluctuations in network properties tend to be more pronounced at smaller scales.

Through an extensive series of experiments, we have meticulously tracked the behavior of SBC as networks evolve under the ER model. We have found that the specific node count at which these peaks occur is not deterministic, introducing an element of unpredictability into the system. Given the inherent variability and uncertainty in peak occurrence, relying solely on a large dataset to discern a definitive pattern becomes impractical. Instead, we have adopted a randomized approach to experiment selection, which allows us to explore a diverse range of network configurations and gain a more comprehensive understanding of the underlying phenomena. This approach not only enhances the robustness of our findings but also enables us to uncover subtle patterns and relationships that may be overlooked in more deterministic approaches.

\begin{figure}[htbp]
    \includegraphics[width=8.6cm]{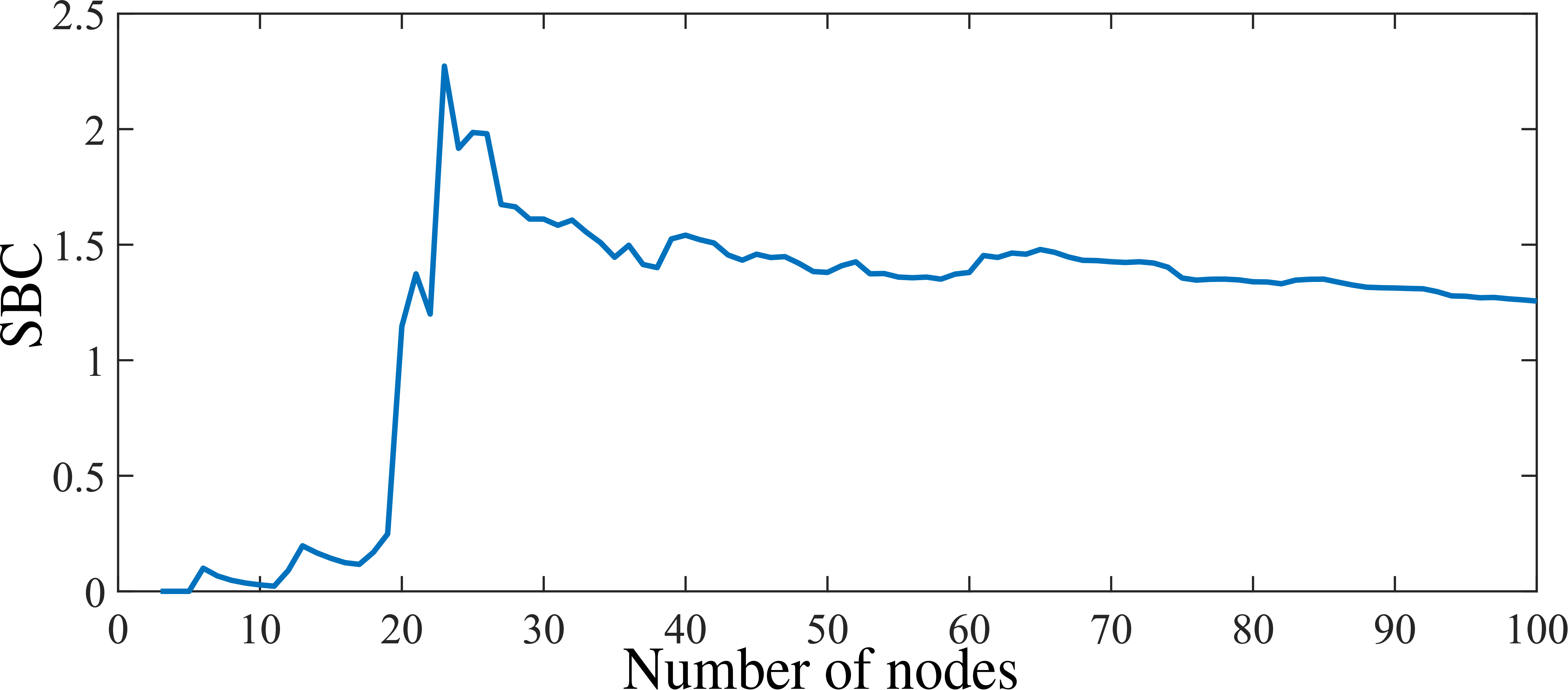}
\caption{The line in the figure shows the change in SBC when the size of the network growing under the ER rule is smaller. It can be found that SBC will have a step increase when some nodes join, and after increasing to a peak value, SBC will gradually decrease and become stable.}

\label{ER-100}
\end{figure}

\begin{figure}[htbp]
    \includegraphics[width=8.6cm]{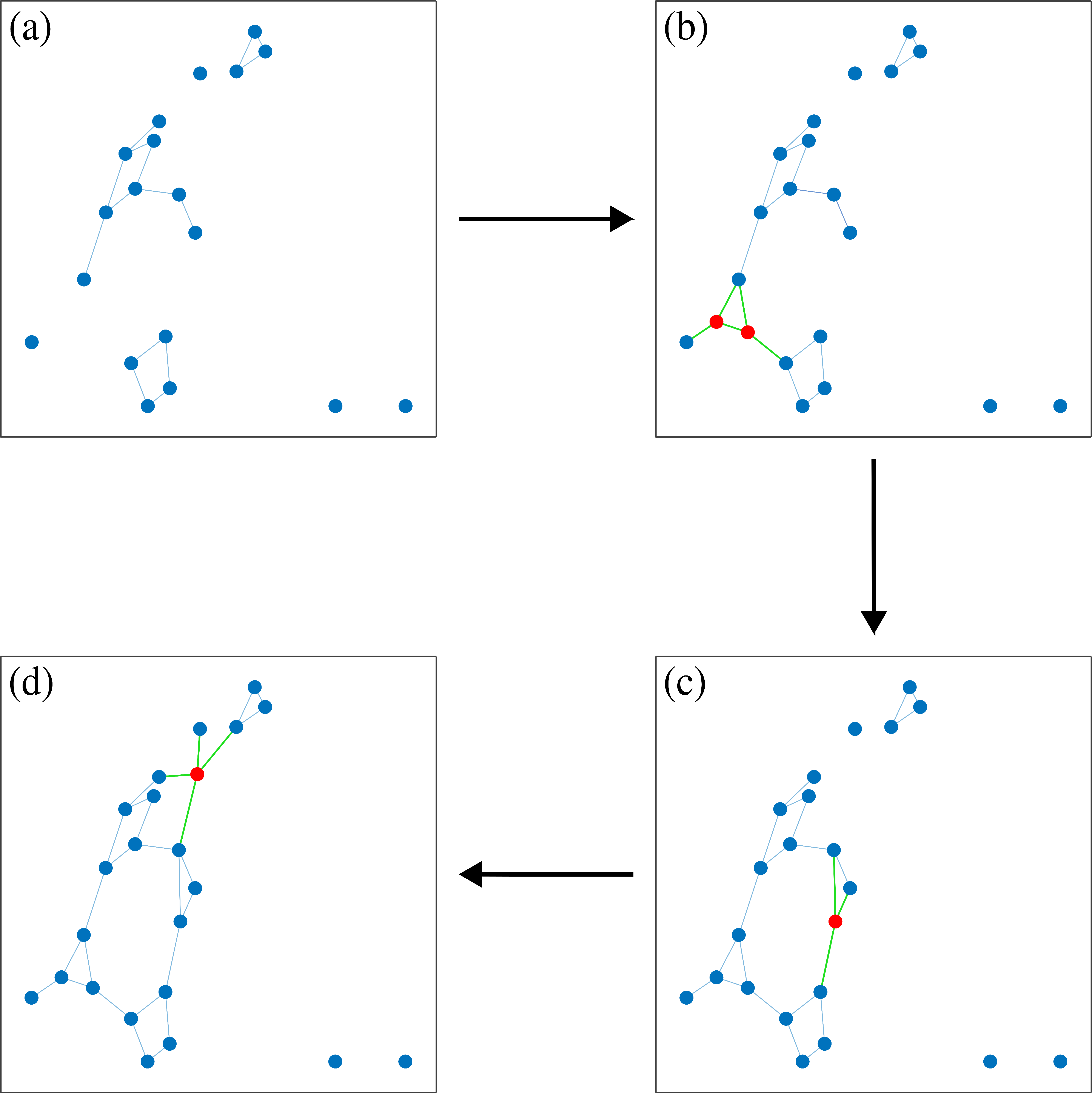}
\caption{The first arrow in the diagram illustrates the initial step depicted in Fig. \ref{ER-100}, while the second arrow demonstrates the subsequent decrease in SBC following this step. The third arrow indicates the occurrence of the peak. In the second network diagram, the red nodes represent the 20th and 21st nodes, while in the third and fourth network diagrams, they represent the 22nd and 23rd nodes, respectively. It is evident that the network begins with numerous isolated parts or nodes, with the initial structure biased towards a core-periphery arrangement. The presence of a few crucial nodes (depicted in red) facilitates the connection of these isolated parts of the network, eventually leading to the integration of all isolated structures.}
\label{network grow}
\end{figure}

Fig.\ref{ER-100} provides the changes in SBC for one such network. However, it is important to note that changes in SBC alone may not provide a complete picture of network dynamics. To gain a deeper understanding, we have also analyzed the structural evolution of the network, focusing on significant alterations that coincide with the attainment of peak SBC values. Fig.\ref{network grow} illustrates these structural changes, providing additional insights into the underlying mechanisms driving peak occurrence. By examining both SBC dynamics and network structure evolution in tandem, we aim to elucidate the factors contributing to these peaks and uncover the underlying mechanisms governing network behaviour.

\begin{figure}[!htbp]
\centering
    \includegraphics[width=8.6cm]{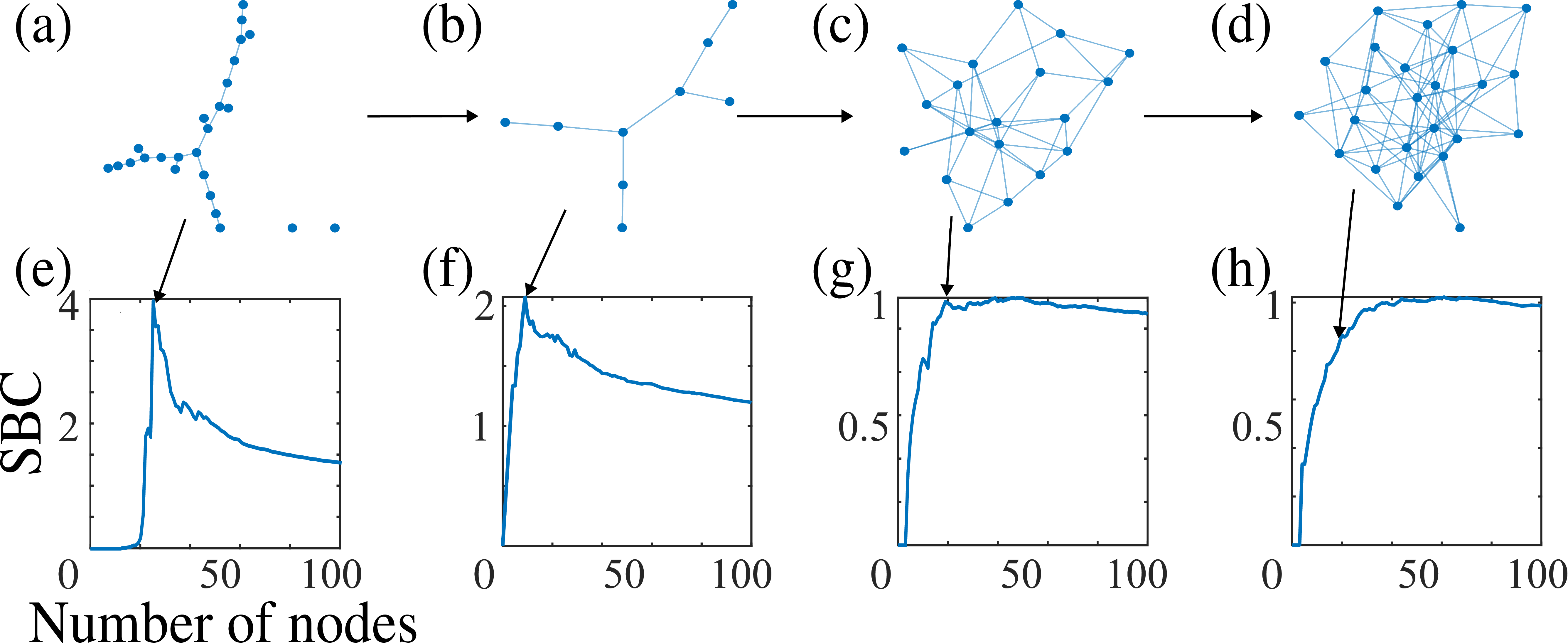}
\caption{The figures display four graphs illustrating how SBC changes with the number of nodes, corresponding to at least 0, 1, 2, and 3 edges from left to right. Additionally, we present network structure graphs highlighting the appearance of the peak. 
}
\label{peak elimination}
\end{figure}

To validate our hypotheses regarding the causes of the observed peaks, we have introduced modifications to the network growth rule aimed at altering the initial network structure to mitigate peak occurrences. Fig.\ref{peak elimination} outlines the process of peak elimination, demonstrating the effectiveness of these modifications in reducing peak occurrence and providing further support for our proposed explanations.

From Fig.\ref{ER-100}, it is evident that the evolution of the SBC aligns with the observations in Fig.\ref{ER}. Furthermore, it is notable that multiple peak values exist. Following the addition of new nodes, exemplified by the 20th, 21st, and 23rd nodes in this experiment, SBC experiences a step increase culminating in the peak value. Conversely, the introduction of certain nodes, such as the 22nd node, leads to a decrease in SBC. Consequently, it becomes imperative to analyze the structural changes within the network during this process in detail. Fig.\ref{network grow} illustrates the dynamics of network structure alterations throughout this progression.

The initial network structure displays significant fragmentation, divided into three main components with numerous isolated nodes. The betweenness centrality within these isolated segments is extremely low; for example, in a triangular network, each node's betweenness centrality is 0, indicating that isolated nodes also have 0 betweenness centrality.

Upon examining Fig. \ref{ER-100} and Fig. \ref{network grow}, it's clear that the initial network's sum of betweenness centralities (SBC) is notably low. The addition of the 20th and 21st nodes in the initial phase connects two isolated segments and one isolated node, resulting in a significant increase in SBC. Essentially, nodes with minimal or 0 betweenness centrality become integrated into another network structure, driving rapid SBC growth.

Due to our growth rule, networks under the ER rule have fewer nodes and links generated initially, favoring a core-periphery structure and rapid SBC growth. However, as our rule favors ER, some nodes make the network structure more uniform. The appearance of the 22nd node leads to a slight decline in SBC. Additionally, isolated nodes contribute to this decline.

As most network segments and isolated nodes integrate, SBC reaches its maximum value. Further node additions induce insignificant SBC changes, albeit possible with low probability. The network increasingly favors growth under the ER rule, leading to gradual SBC stabilization. Thus, SBC reflects network structure changes even with a small network scale.

We believe the peak occurs because our ER growth rule biases the initial network structure towards a core-periphery setup. It's akin to initially using the BA rule for network growth and then transitioning to the ER rule, leading to peak emergence. To validate this, we conduct four experiments, each employing ER rules with new nodes generating at least 0, 1, 2, or 3 edges.Fig.\ref{peak elimination} shows our peak elimination process.
\nopagebreak

Notably, after generating at least one edge, the peak diminishes and occurs earlier. This phenomenon arises because the network structure gradually shifts towards a core-peripheral configuration, resulting in fewer nodes experiencing such peaks and a quicker transition towards network homogeneity. Moreover, the second graph's network diagram reveals a smaller number of nodes where the peak occurs, leading to a lower peak value. Subsequently, as at least two or three edges are generated, the latter two graphs demonstrate that with more edges in the network, the network becomes more uniform, causing the peaks to decrease and eventually vanish. This observation confirms our suspicion that the spike is attributed to a change in the network structure, thereby enhancing our understanding of SBC.

\section{Sum of nodes' betweenness centrality in real network}

Merely real network data isn't sufficient to investigate the variation of SBC in the growth of real networks. Hence, we require a method to systematically remove nodes from real networks to observe the SBC change. Thus, we employed the K-shell approach to gradually downsize the real networks. The procedure of K-shell implementation is as follows: Initially, nodes and edges with a degree of 1 are deleted from the network, followed by nodes and edges with a degree of 2, iteratively continuing until all nodes are removed.

One advantage of this method is that the node removal sequence aligns closely with the node addition sequence under different rules. Even if not identical, early removed nodes have minimal impact on the network structure. For instance, in networks grown under the BA rule, the preference for deleting one of the two degree-1 nodes doesn't significantly affect SBC.

\begin{figure}[htbp]
\centering
    \includegraphics[width=8.6cm]{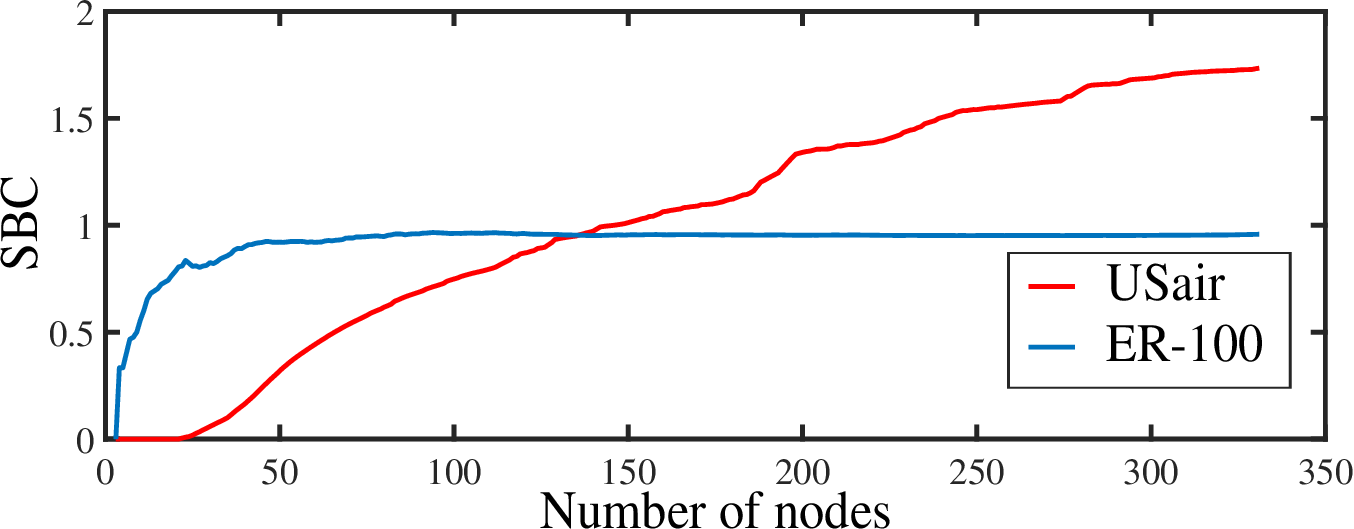}
\caption{
The red line in the figure shows the change of SBC obtained after processing US-Airline using K-shell method, which is distributed exponentially. The blue line is a network of 332 nodes that we grow using ER growth rules, and the K-shell method is used to process it. It can be seen that SBC is stable. 
}
\label{USair}
\end{figure}

Selecting real networks biased towards structures growing under BA or ER rules is ideal. However, real-world networks often lack homogeneity. Hence, we utilized our own ER rule-grown network and a core-periphery structured network, specifically the US-Airline network~(Pajek, 2006). The K-shell method was applied to remove nodes from these two networks, capturing the corresponding SBC changes. This process is illustrated in Fig. \ref{USair}.

The SBC in the USair network aligns with the scenario depicted in Figure 1, indicating that USair exhibits a core-periphery structure, as depicted. This underscores the practicality of the SBC proposed in this paper for real-world networks, showcasing its strong reliability across both simulation experiments and real-world applications. Notably, in the figure, the peak value of the network, growing to 332 nodes under the ER rule, diminishes upon employing the K-shell method. The K-shell method prioritizes the removal of nodes with low degrees, yet crucial nodes driving substantial SBC growth in the network's early stages remain unaffected due to their connections with isolated network structures and nodes, resulting in the disappearance of the peak. This principle echoes the peak elimination concept illustrated in Figure 6, aimed at homogenizing the initial network structure to eliminate peaks. Consequently, this reverse experiment further elucidates the sensitivity of SBC to network structure.

\section{Conclusions}
The SBC is an indicator capable of reflecting changes in network structure.
In the network's growth guided by the BA rule, SBC exhibits logarithmic growth, which leads to the core-periphery structure. Concurrently, as core nodes' degree and betweenness steadily rise with node additions, the deceleration in SBC growth symbolizes the heightened centrality of select nodes within the network. Conversely, the network's growth under ER rules demonstrates a distinct stable state in SBC. Here, adding new nodes preserves the network's uniform structure, ensuring the enduring stability of SBC.

The SBC illuminates the network's structural evolution. Altering growth rules mid-process significantly impacts SBC, providing a more intuitive measure of the network's growth rule and a quicker and more direct insight into these alterations.

Surprisingly, in smaller networks, SBC shows a unique phenomenon: its fluctuations are connected with the changes in the network's connectivity. The rapid SBC growth signifies the integration of isolated network segments with the addition of new nodes, diverging from core-periphery structured networks, thus shedding light on nodes' significance from an alternate perspective.

In summary, SBC can be used as a new index for structural characteristics, which shows the network's underlying growth rules.

\section*{Acknowledgments}
This work is supported by the Scientific Research Funding of Jiangsu University of Science and Technology (No.1052932204), the National Natural Science Foundation of China (Grant No. 62303198), the Research Initiation Fund for Senior Talents of Jiangsu University (No. 5501170008).

\bibliography{ref}

%merlin.mbs apsrev4-1.bst 2010-07-25 4.21a (PWD, AO, DPC) hacked
%Control: key (0)
%Control: author (72) initials jnrlst
%Control: editor formatted (1) identically to author
%Control: production of article title (-1) disabled
%Control: page (0) single
%Control: year (1) truncated
%Control: production of eprint (0) enabled
\begin{thebibliography}{30}%
\makeatletter
\providecommand \@ifxundefined [1]{%
 \@ifx{#1\undefined}
}%
\providecommand \@ifnum [1]{%
 \ifnum #1\expandafter \@firstoftwo
 \else \expandafter \@secondoftwo
 \fi
}%
\providecommand \@ifx [1]{%
 \ifx #1\expandafter \@firstoftwo
 \else \expandafter \@secondoftwo
 \fi
}%
\providecommand \natexlab [1]{#1}%
\providecommand \enquote  [1]{``#1''}%
\providecommand \bibnamefont  [1]{#1}%
\providecommand \bibfnamefont [1]{#1}%
\providecommand \citenamefont [1]{#1}%
\providecommand \href@noop [0]{\@secondoftwo}%
\providecommand \href [0]{\begingroup \@sanitize@url \@href}%
\providecommand \@href[1]{\@@startlink{#1}\@@href}%
\providecommand \@@href[1]{\endgroup#1\@@endlink}%
\providecommand \@sanitize@url [0]{\catcode `\\12\catcode `\$12\catcode
  `\&12\catcode `\#12\catcode `\^12\catcode `\_12\catcode `\%12\relax}%
\providecommand \@@startlink[1]{}%
\providecommand \@@endlink[0]{}%
\providecommand \url  [0]{\begingroup\@sanitize@url \@url }%
\providecommand \@url [1]{\endgroup\@href {#1}{\urlprefix }}%
\providecommand \urlprefix  [0]{URL }%
\providecommand \Eprint [0]{\href }%
\providecommand \doibase [0]{http://dx.doi.org/}%
\providecommand \selectlanguage [0]{\@gobble}%
\providecommand \bibinfo  [0]{\@secondoftwo}%
\providecommand \bibfield  [0]{\@secondoftwo}%
\providecommand \translation [1]{[#1]}%
\providecommand \BibitemOpen [0]{}%
\providecommand \bibitemStop [0]{}%
\providecommand \bibitemNoStop [0]{.\EOS\space}%
\providecommand \EOS [0]{\spacefactor3000\relax}%
\providecommand \BibitemShut  [1]{\csname bibitem#1\endcsname}%
\let\auto@bib@innerbib\@empty
%</preamble>
\bibitem [{\citenamefont {Watts}\ and\ \citenamefont
  {Strogatz}(1998)}]{watts1998collective}%
  \BibitemOpen
  \bibfield  {author} {\bibinfo {author} {\bibfnamefont {D.~J.}\ \bibnamefont
  {Watts}}\ and\ \bibinfo {author} {\bibfnamefont {S.~H.}\ \bibnamefont
  {Strogatz}},\ }\href@noop {} {\bibfield  {journal} {\bibinfo  {journal}
  {nature}\ }\textbf {\bibinfo {volume} {393}},\ \bibinfo {pages} {440}
  (\bibinfo {year} {1998})}\BibitemShut {NoStop}%
\bibitem [{\citenamefont {Barab{\'a}si}\ and\ \citenamefont
  {Bonabeau}(2003)}]{barabasi2003scale}%
  \BibitemOpen
  \bibfield  {author} {\bibinfo {author} {\bibfnamefont {A.-L.}\ \bibnamefont
  {Barab{\'a}si}}\ and\ \bibinfo {author} {\bibfnamefont {E.}~\bibnamefont
  {Bonabeau}},\ }\href@noop {} {\bibfield  {journal} {\bibinfo  {journal}
  {Scientific american}\ }\textbf {\bibinfo {volume} {288}},\ \bibinfo {pages}
  {60} (\bibinfo {year} {2003})}\BibitemShut {NoStop}%
\bibitem [{\citenamefont {Newman}(2004)}]{newman2004analysis}%
  \BibitemOpen
  \bibfield  {author} {\bibinfo {author} {\bibfnamefont {M.~E.}\ \bibnamefont
  {Newman}},\ }\href@noop {} {\bibfield  {journal} {\bibinfo  {journal}
  {Physical review E}\ }\textbf {\bibinfo {volume} {70}},\ \bibinfo {pages}
  {056131} (\bibinfo {year} {2004})}\BibitemShut {NoStop}%
\bibitem [{\citenamefont {Garlaschelli}\ and\ \citenamefont
  {Loffredo}(2009)}]{garlaschelli2009generalized}%
  \BibitemOpen
  \bibfield  {author} {\bibinfo {author} {\bibfnamefont {D.}~\bibnamefont
  {Garlaschelli}}\ and\ \bibinfo {author} {\bibfnamefont {M.~I.}\ \bibnamefont
  {Loffredo}},\ }\href@noop {} {\bibfield  {journal} {\bibinfo  {journal}
  {Physical review letters}\ }\textbf {\bibinfo {volume} {102}},\ \bibinfo
  {pages} {038701} (\bibinfo {year} {2009})}\BibitemShut {NoStop}%
\bibitem [{\citenamefont {Bianconi}\ and\ \citenamefont
  {Barab{\'a}si}(2001)}]{bianconi2001bose}%
  \BibitemOpen
  \bibfield  {author} {\bibinfo {author} {\bibfnamefont {G.}~\bibnamefont
  {Bianconi}}\ and\ \bibinfo {author} {\bibfnamefont {A.-L.}\ \bibnamefont
  {Barab{\'a}si}},\ }\href@noop {} {\bibfield  {journal} {\bibinfo  {journal}
  {Physical review letters}\ }\textbf {\bibinfo {volume} {86}},\ \bibinfo
  {pages} {5632} (\bibinfo {year} {2001})}\BibitemShut {NoStop}%
\bibitem [{\citenamefont {Von~Mering}\ \emph {et~al.}(2002)\citenamefont
  {Von~Mering}, \citenamefont {Krause}, \citenamefont {Snel}, \citenamefont
  {Cornell}, \citenamefont {Oliver}, \citenamefont {Fields},\ and\
  \citenamefont {Bork}}]{von2002comparative}%
  \BibitemOpen
  \bibfield  {author} {\bibinfo {author} {\bibfnamefont {C.}~\bibnamefont
  {Von~Mering}}, \bibinfo {author} {\bibfnamefont {R.}~\bibnamefont {Krause}},
  \bibinfo {author} {\bibfnamefont {B.}~\bibnamefont {Snel}}, \bibinfo {author}
  {\bibfnamefont {M.}~\bibnamefont {Cornell}}, \bibinfo {author} {\bibfnamefont
  {S.~G.}\ \bibnamefont {Oliver}}, \bibinfo {author} {\bibfnamefont
  {S.}~\bibnamefont {Fields}}, \ and\ \bibinfo {author} {\bibfnamefont
  {P.}~\bibnamefont {Bork}},\ }\href@noop {} {\bibfield  {journal} {\bibinfo
  {journal} {Nature}\ }\textbf {\bibinfo {volume} {417}},\ \bibinfo {pages}
  {399} (\bibinfo {year} {2002})}\BibitemShut {NoStop}%
\bibitem [{\citenamefont {Korcsmaros}\ \emph {et~al.}(2017)\citenamefont
  {Korcsmaros}, \citenamefont {Schneider},\ and\ \citenamefont
  {Superti-Furga}}]{korcsmaros2017next}%
  \BibitemOpen
  \bibfield  {author} {\bibinfo {author} {\bibfnamefont {T.}~\bibnamefont
  {Korcsmaros}}, \bibinfo {author} {\bibfnamefont {M.~V.}\ \bibnamefont
  {Schneider}}, \ and\ \bibinfo {author} {\bibfnamefont {G.}~\bibnamefont
  {Superti-Furga}},\ }\href@noop {} {\bibfield  {journal} {\bibinfo  {journal}
  {Integrative Biology}\ }\textbf {\bibinfo {volume} {9}},\ \bibinfo {pages}
  {97} (\bibinfo {year} {2017})}\BibitemShut {NoStop}%
\bibitem [{\citenamefont {Gosak}\ \emph {et~al.}(2018)\citenamefont {Gosak},
  \citenamefont {Markovi{\v{c}}}, \citenamefont {Dolen{\v{s}}ek}, \citenamefont
  {Rupnik}, \citenamefont {Marhl}, \citenamefont {Sto{\v{z}}er},\ and\
  \citenamefont {Perc}}]{gosak2018network}%
  \BibitemOpen
  \bibfield  {author} {\bibinfo {author} {\bibfnamefont {M.}~\bibnamefont
  {Gosak}}, \bibinfo {author} {\bibfnamefont {R.}~\bibnamefont
  {Markovi{\v{c}}}}, \bibinfo {author} {\bibfnamefont {J.}~\bibnamefont
  {Dolen{\v{s}}ek}}, \bibinfo {author} {\bibfnamefont {M.~S.}\ \bibnamefont
  {Rupnik}}, \bibinfo {author} {\bibfnamefont {M.}~\bibnamefont {Marhl}},
  \bibinfo {author} {\bibfnamefont {A.}~\bibnamefont {Sto{\v{z}}er}}, \ and\
  \bibinfo {author} {\bibfnamefont {M.}~\bibnamefont {Perc}},\ }\href@noop {}
  {\bibfield  {journal} {\bibinfo  {journal} {Physics of life reviews}\
  }\textbf {\bibinfo {volume} {24}},\ \bibinfo {pages} {118} (\bibinfo {year}
  {2018})}\BibitemShut {NoStop}%
\bibitem [{\citenamefont {Shai}\ \emph {et~al.}(2017)\citenamefont {Shai},
  \citenamefont {Porter}, \citenamefont {Pascual},\ and\ \citenamefont
  {Sonia}}]{shai2017multilayer}%
  \BibitemOpen
  \bibfield  {author} {\bibinfo {author} {\bibfnamefont {P.}~\bibnamefont
  {Shai}}, \bibinfo {author} {\bibfnamefont {M.~A.}\ \bibnamefont {Porter}},
  \bibinfo {author} {\bibfnamefont {M.}~\bibnamefont {Pascual}}, \ and\
  \bibinfo {author} {\bibfnamefont {K.}~\bibnamefont {Sonia}},\ }\href@noop {}
  {\bibfield  {journal} {\bibinfo  {journal} {Nature Ecology \& Evolution}\
  }\textbf {\bibinfo {volume} {1}} (\bibinfo {year} {2017})}\BibitemShut
  {NoStop}%
\bibitem [{\citenamefont {Schneidman}\ \emph {et~al.}(2006)\citenamefont
  {Schneidman}, \citenamefont {Berry}, \citenamefont {Segev},\ and\
  \citenamefont {Bialek}}]{schneidman2006weak}%
  \BibitemOpen
  \bibfield  {author} {\bibinfo {author} {\bibfnamefont {E.}~\bibnamefont
  {Schneidman}}, \bibinfo {author} {\bibfnamefont {M.~J.}\ \bibnamefont
  {Berry}}, \bibinfo {author} {\bibfnamefont {R.}~\bibnamefont {Segev}}, \ and\
  \bibinfo {author} {\bibfnamefont {W.}~\bibnamefont {Bialek}},\ }\href@noop {}
  {\bibfield  {journal} {\bibinfo  {journal} {Nature}\ }\textbf {\bibinfo
  {volume} {440}},\ \bibinfo {pages} {1007} (\bibinfo {year}
  {2006})}\BibitemShut {NoStop}%
\bibitem [{\citenamefont {Bassett}\ and\ \citenamefont
  {Sporns}(2017)}]{bassett2017network}%
  \BibitemOpen
  \bibfield  {author} {\bibinfo {author} {\bibfnamefont {D.~S.}\ \bibnamefont
  {Bassett}}\ and\ \bibinfo {author} {\bibfnamefont {O.}~\bibnamefont
  {Sporns}},\ }\href@noop {} {\bibfield  {journal} {\bibinfo  {journal} {Nature
  neuroscience}\ }\textbf {\bibinfo {volume} {20}},\ \bibinfo {pages} {353}
  (\bibinfo {year} {2017})}\BibitemShut {NoStop}%
\bibitem [{\citenamefont {Battiston}\ \emph {et~al.}(2016)\citenamefont
  {Battiston}, \citenamefont {Farmer}, \citenamefont {Flache}, \citenamefont
  {Garlaschelli}, \citenamefont {Haldane}, \citenamefont {Heesterbeek},
  \citenamefont {Hommes}, \citenamefont {Jaeger}, \citenamefont {May},\ and\
  \citenamefont {Scheffer}}]{battiston2016complexity}%
  \BibitemOpen
  \bibfield  {author} {\bibinfo {author} {\bibfnamefont {S.}~\bibnamefont
  {Battiston}}, \bibinfo {author} {\bibfnamefont {J.~D.}\ \bibnamefont
  {Farmer}}, \bibinfo {author} {\bibfnamefont {A.}~\bibnamefont {Flache}},
  \bibinfo {author} {\bibfnamefont {D.}~\bibnamefont {Garlaschelli}}, \bibinfo
  {author} {\bibfnamefont {A.~G.}\ \bibnamefont {Haldane}}, \bibinfo {author}
  {\bibfnamefont {H.}~\bibnamefont {Heesterbeek}}, \bibinfo {author}
  {\bibfnamefont {C.}~\bibnamefont {Hommes}}, \bibinfo {author} {\bibfnamefont
  {C.}~\bibnamefont {Jaeger}}, \bibinfo {author} {\bibfnamefont
  {R.}~\bibnamefont {May}}, \ and\ \bibinfo {author} {\bibfnamefont
  {M.}~\bibnamefont {Scheffer}},\ }\href@noop {} {\bibfield  {journal}
  {\bibinfo  {journal} {Science}\ }\textbf {\bibinfo {volume} {351}},\ \bibinfo
  {pages} {818} (\bibinfo {year} {2016})}\BibitemShut {NoStop}%
\bibitem [{\citenamefont {Almog}\ and\ \citenamefont
  {Garlaschelli}(2014)}]{almog2014binary}%
  \BibitemOpen
  \bibfield  {author} {\bibinfo {author} {\bibfnamefont {A.}~\bibnamefont
  {Almog}}\ and\ \bibinfo {author} {\bibfnamefont {D.}~\bibnamefont
  {Garlaschelli}},\ }\href@noop {} {\bibfield  {journal} {\bibinfo  {journal}
  {New journal of physics}\ }\textbf {\bibinfo {volume} {16}},\ \bibinfo
  {pages} {093015} (\bibinfo {year} {2014})}\BibitemShut {NoStop}%
\bibitem [{\citenamefont {Kenett}\ and\ \citenamefont
  {Havlin}(2015)}]{kenett2015network}%
  \BibitemOpen
  \bibfield  {author} {\bibinfo {author} {\bibfnamefont {D.~Y.}\ \bibnamefont
  {Kenett}}\ and\ \bibinfo {author} {\bibfnamefont {S.}~\bibnamefont
  {Havlin}},\ }\href@noop {} {\bibfield  {journal} {\bibinfo  {journal} {Mind
  \& Society}\ }\textbf {\bibinfo {volume} {14}},\ \bibinfo {pages} {155}
  (\bibinfo {year} {2015})}\BibitemShut {NoStop}%
\bibitem [{\citenamefont {Bello-Orgaz}\ \emph {et~al.}(2016)\citenamefont
  {Bello-Orgaz}, \citenamefont {Jung},\ and\ \citenamefont
  {Camacho}}]{bello2016social}%
  \BibitemOpen
  \bibfield  {author} {\bibinfo {author} {\bibfnamefont {G.}~\bibnamefont
  {Bello-Orgaz}}, \bibinfo {author} {\bibfnamefont {J.~J.}\ \bibnamefont
  {Jung}}, \ and\ \bibinfo {author} {\bibfnamefont {D.}~\bibnamefont
  {Camacho}},\ }\href@noop {} {\bibfield  {journal} {\bibinfo  {journal}
  {Information Fusion}\ }\textbf {\bibinfo {volume} {28}},\ \bibinfo {pages}
  {45} (\bibinfo {year} {2016})}\BibitemShut {NoStop}%
\bibitem [{\citenamefont {Jusup}\ \emph {et~al.}(2022)\citenamefont {Jusup},
  \citenamefont {Holme}, \citenamefont {Kanazawa}, \citenamefont {Takayasu},
  \citenamefont {Romi{\'c}}, \citenamefont {Wang}, \citenamefont {Ge{\v{c}}ek},
  \citenamefont {Lipi{\'c}}, \citenamefont {Podobnik}, \citenamefont {Wang}
  \emph {et~al.}}]{jusup2022social}%
  \BibitemOpen
  \bibfield  {author} {\bibinfo {author} {\bibfnamefont {M.}~\bibnamefont
  {Jusup}}, \bibinfo {author} {\bibfnamefont {P.}~\bibnamefont {Holme}},
  \bibinfo {author} {\bibfnamefont {K.}~\bibnamefont {Kanazawa}}, \bibinfo
  {author} {\bibfnamefont {M.}~\bibnamefont {Takayasu}}, \bibinfo {author}
  {\bibfnamefont {I.}~\bibnamefont {Romi{\'c}}}, \bibinfo {author}
  {\bibfnamefont {Z.}~\bibnamefont {Wang}}, \bibinfo {author} {\bibfnamefont
  {S.}~\bibnamefont {Ge{\v{c}}ek}}, \bibinfo {author} {\bibfnamefont
  {T.}~\bibnamefont {Lipi{\'c}}}, \bibinfo {author} {\bibfnamefont
  {B.}~\bibnamefont {Podobnik}}, \bibinfo {author} {\bibfnamefont
  {L.}~\bibnamefont {Wang}},  \emph {et~al.},\ }\href@noop {} {\bibfield
  {journal} {\bibinfo  {journal} {Physics Reports}\ }\textbf {\bibinfo {volume}
  {948}},\ \bibinfo {pages} {1} (\bibinfo {year} {2022})}\BibitemShut {NoStop}%
\bibitem [{\citenamefont {Cimini}\ \emph {et~al.}(2019)\citenamefont {Cimini},
  \citenamefont {Squartini}, \citenamefont {Saracco}, \citenamefont
  {Garlaschelli}, \citenamefont {Gabrielli},\ and\ \citenamefont
  {Caldarelli}}]{cimini2019statistical}%
  \BibitemOpen
  \bibfield  {author} {\bibinfo {author} {\bibfnamefont {G.}~\bibnamefont
  {Cimini}}, \bibinfo {author} {\bibfnamefont {T.}~\bibnamefont {Squartini}},
  \bibinfo {author} {\bibfnamefont {F.}~\bibnamefont {Saracco}}, \bibinfo
  {author} {\bibfnamefont {D.}~\bibnamefont {Garlaschelli}}, \bibinfo {author}
  {\bibfnamefont {A.}~\bibnamefont {Gabrielli}}, \ and\ \bibinfo {author}
  {\bibfnamefont {G.}~\bibnamefont {Caldarelli}},\ }\href@noop {} {\bibfield
  {journal} {\bibinfo  {journal} {Nature Reviews Physics}\ }\textbf {\bibinfo
  {volume} {1}},\ \bibinfo {pages} {58} (\bibinfo {year} {2019})}\BibitemShut
  {NoStop}%
\bibitem [{\citenamefont {Battiston}\ \emph {et~al.}(2020)\citenamefont
  {Battiston}, \citenamefont {Cencetti}, \citenamefont {Iacopini},
  \citenamefont {Latora}, \citenamefont {Lucas}, \citenamefont {Patania},
  \citenamefont {Young},\ and\ \citenamefont {Petri}}]{battiston2020networks}%
  \BibitemOpen
  \bibfield  {author} {\bibinfo {author} {\bibfnamefont {F.}~\bibnamefont
  {Battiston}}, \bibinfo {author} {\bibfnamefont {G.}~\bibnamefont {Cencetti}},
  \bibinfo {author} {\bibfnamefont {I.}~\bibnamefont {Iacopini}}, \bibinfo
  {author} {\bibfnamefont {V.}~\bibnamefont {Latora}}, \bibinfo {author}
  {\bibfnamefont {M.}~\bibnamefont {Lucas}}, \bibinfo {author} {\bibfnamefont
  {A.}~\bibnamefont {Patania}}, \bibinfo {author} {\bibfnamefont {J.-G.}\
  \bibnamefont {Young}}, \ and\ \bibinfo {author} {\bibfnamefont
  {G.}~\bibnamefont {Petri}},\ }\href@noop {} {\bibfield  {journal} {\bibinfo
  {journal} {Physics Reports}\ }\textbf {\bibinfo {volume} {874}},\ \bibinfo
  {pages} {1} (\bibinfo {year} {2020})}\BibitemShut {NoStop}%
\bibitem [{\citenamefont {Squartini}\ \emph {et~al.}(2015)\citenamefont
  {Squartini}, \citenamefont {de~Mol}, \citenamefont {den Hollander},\ and\
  \citenamefont {Garlaschelli}}]{squartini2015breaking}%
  \BibitemOpen
  \bibfield  {author} {\bibinfo {author} {\bibfnamefont {T.}~\bibnamefont
  {Squartini}}, \bibinfo {author} {\bibfnamefont {J.}~\bibnamefont {de~Mol}},
  \bibinfo {author} {\bibfnamefont {F.}~\bibnamefont {den Hollander}}, \ and\
  \bibinfo {author} {\bibfnamefont {D.}~\bibnamefont {Garlaschelli}},\
  }\href@noop {} {\bibfield  {journal} {\bibinfo  {journal} {Physical review
  letters}\ }\textbf {\bibinfo {volume} {115}},\ \bibinfo {pages} {268701}
  (\bibinfo {year} {2015})}\BibitemShut {NoStop}%
\bibitem [{\citenamefont {Barvinok}(2012)}]{barvinok2012matrices}%
  \BibitemOpen
  \bibfield  {author} {\bibinfo {author} {\bibfnamefont {A.}~\bibnamefont
  {Barvinok}},\ }\href@noop {} {\bibfield  {journal} {\bibinfo  {journal}
  {Linear Algebra and its Applications}\ }\textbf {\bibinfo {volume} {436}},\
  \bibinfo {pages} {820} (\bibinfo {year} {2012})}\BibitemShut {NoStop}%
\bibitem [{\citenamefont {Hollander}\ \emph {et~al.}(2018)\citenamefont
  {Hollander}, \citenamefont {Mandjes}, \citenamefont {Roccaverde},\ and\
  \citenamefont {Starreveld}}]{hollander2018breaking}%
  \BibitemOpen
  \bibfield  {author} {\bibinfo {author} {\bibfnamefont {F.~d.}\ \bibnamefont
  {Hollander}}, \bibinfo {author} {\bibfnamefont {M.}~\bibnamefont {Mandjes}},
  \bibinfo {author} {\bibfnamefont {A.}~\bibnamefont {Roccaverde}}, \ and\
  \bibinfo {author} {\bibfnamefont {N.}~\bibnamefont {Starreveld}},\
  }\href@noop {} {\bibfield  {journal} {\bibinfo  {journal} {arXiv preprint
  arXiv:1807.07750}\ } (\bibinfo {year} {2018})}\BibitemShut {NoStop}%
\bibitem [{\citenamefont {Squartini}\ and\ \citenamefont
  {Garlaschelli}(2017)}]{squartini2017reconnecting}%
  \BibitemOpen
  \bibfield  {author} {\bibinfo {author} {\bibfnamefont {T.}~\bibnamefont
  {Squartini}}\ and\ \bibinfo {author} {\bibfnamefont {D.}~\bibnamefont
  {Garlaschelli}},\ }\href@noop {} {\bibfield  {journal} {\bibinfo  {journal}
  {arXiv preprint arXiv:1710.11422}\ } (\bibinfo {year} {2017})}\BibitemShut
  {NoStop}%
\bibitem [{\citenamefont {Zhang}\ and\ \citenamefont
  {Garlaschelli}(2022)}]{zhang2022strong}%
  \BibitemOpen
  \bibfield  {author} {\bibinfo {author} {\bibfnamefont {Q.}~\bibnamefont
  {Zhang}}\ and\ \bibinfo {author} {\bibfnamefont {D.}~\bibnamefont
  {Garlaschelli}},\ }\href@noop {} {\bibfield  {journal} {\bibinfo  {journal}
  {New Journal of Physics}\ }\textbf {\bibinfo {volume} {24}},\ \bibinfo
  {pages} {043011} (\bibinfo {year} {2022})}\BibitemShut {NoStop}%
\bibitem [{\citenamefont {Prountzos}\ and\ \citenamefont
  {Pingali}(2013)}]{prountzos2013betweenness}%
  \BibitemOpen
  \bibfield  {author} {\bibinfo {author} {\bibfnamefont {D.}~\bibnamefont
  {Prountzos}}\ and\ \bibinfo {author} {\bibfnamefont {K.}~\bibnamefont
  {Pingali}},\ }in\ \href@noop {} {\emph {\bibinfo {booktitle} {Proceedings of
  the 18th ACM SIGPLAN symposium on Principles and practice of parallel
  programming}}}\ (\bibinfo {year} {2013})\ pp.\ \bibinfo {pages}
  {35--46}\BibitemShut {NoStop}%
\bibitem [{\citenamefont {Lee}\ \emph {et~al.}(2012)\citenamefont {Lee},
  \citenamefont {Lee}, \citenamefont {Park}, \citenamefont {Choi},\ and\
  \citenamefont {Chung}}]{lee2012qube}%
  \BibitemOpen
  \bibfield  {author} {\bibinfo {author} {\bibfnamefont {M.-J.}\ \bibnamefont
  {Lee}}, \bibinfo {author} {\bibfnamefont {J.}~\bibnamefont {Lee}}, \bibinfo
  {author} {\bibfnamefont {J.~Y.}\ \bibnamefont {Park}}, \bibinfo {author}
  {\bibfnamefont {R.~H.}\ \bibnamefont {Choi}}, \ and\ \bibinfo {author}
  {\bibfnamefont {C.-W.}\ \bibnamefont {Chung}},\ }in\ \href@noop {} {\emph
  {\bibinfo {booktitle} {Proceedings of the 21st international conference on
  World Wide Web}}}\ (\bibinfo {year} {2012})\ pp.\ \bibinfo {pages}
  {351--360}\BibitemShut {NoStop}%
\bibitem [{\citenamefont {Brandes}(2001)}]{brandes2001faster}%
  \BibitemOpen
  \bibfield  {author} {\bibinfo {author} {\bibfnamefont {U.}~\bibnamefont
  {Brandes}},\ }\href@noop {} {\bibfield  {journal} {\bibinfo  {journal}
  {Journal of mathematical sociology}\ }\textbf {\bibinfo {volume} {25}},\
  \bibinfo {pages} {163} (\bibinfo {year} {2001})}\BibitemShut {NoStop}%
\bibitem [{\citenamefont {Barthelemy}(2004)}]{barthelemy2004betweenness}%
  \BibitemOpen
  \bibfield  {author} {\bibinfo {author} {\bibfnamefont {M.}~\bibnamefont
  {Barthelemy}},\ }\href@noop {} {\bibfield  {journal} {\bibinfo  {journal}
  {The European physical journal B}\ }\textbf {\bibinfo {volume} {38}},\
  \bibinfo {pages} {163} (\bibinfo {year} {2004})}\BibitemShut {NoStop}%
\bibitem [{\citenamefont {Newman}(2005)}]{newman2005measure}%
  \BibitemOpen
  \bibfield  {author} {\bibinfo {author} {\bibfnamefont {M.~E.}\ \bibnamefont
  {Newman}},\ }\href@noop {} {\bibfield  {journal} {\bibinfo  {journal} {Social
  networks}\ }\textbf {\bibinfo {volume} {27}},\ \bibinfo {pages} {39}
  (\bibinfo {year} {2005})}\BibitemShut {NoStop}%
\bibitem [{\citenamefont {Leydesdorff}(2007)}]{leydesdorff2007betweenness}%
  \BibitemOpen
  \bibfield  {author} {\bibinfo {author} {\bibfnamefont {L.}~\bibnamefont
  {Leydesdorff}},\ }\href@noop {} {\bibfield  {journal} {\bibinfo  {journal}
  {Journal of the American Society for Information Science and Technology}\
  }\textbf {\bibinfo {volume} {58}},\ \bibinfo {pages} {1303} (\bibinfo {year}
  {2007})}\BibitemShut {NoStop}%
\bibitem [{\citenamefont {Erd{\H{o}}s}\ and\ \citenamefont
  {R{\'e}nyi}(1961)}]{erdHos1961strength}%
  \BibitemOpen
  \bibfield  {author} {\bibinfo {author} {\bibfnamefont {P.}~\bibnamefont
  {Erd{\H{o}}s}}\ and\ \bibinfo {author} {\bibfnamefont {A.}~\bibnamefont
  {R{\'e}nyi}},\ }\href@noop {} {\bibfield  {journal} {\bibinfo  {journal}
  {Acta Mathematica Hungarica}\ }\textbf {\bibinfo {volume} {12}},\ \bibinfo
  {pages} {261} (\bibinfo {year} {1961})}\BibitemShut {NoStop}%
\end{thebibliography}%
%\bibliographystyle{apsrev4-1}
%\appendix
%\section*{Supplementary Information}

\end{document}